\documentclass[]{aa501}
\usepackage{graphics}
\begin{document}

%\thesaurus{4(11.13.2, 11.19.2, 13.18.1, 02.16.2)}
\title{Magnetic fields in the absence of spiral density waves -- NGC~4414}

\author{M. Soida\inst{1}
 \and R. Beck\inst{3}
 \and M. Urbanik\inst{1}
 \and J. Braine\inst{2}}
\institute{Astronomical Observatory, Jagiellonian University, ul. Orla 171, 
PL-30-244 Krak\'ow, Poland
 \and Observatoire de Bordeaux, UMR 5804 CNRS/INSU, B.P. 89,
F-33270 Floirac, France
 \and Max-Planck-Institut f\"ur Radioastronomie, Auf dem H\"ugel 69,
D-53121 Bonn, Germany }
\offprints{M. Soida}
\mail{soida@oa.uj.edu.pl}

\date{Received November 1st, 2001; accepted July 27th, 2002}
%\markboth{M. Soida et al. Spiral magnetic
%fields in absence.. }
%\markboth{M. Soida et al. Spiral magnetic fields in
%absence... }

\titlerunning{ Spiral magnetic fields in NGC~4414}
\authorrunning{M. Soida et al.}

\abstract{
We present three-frequency VLA observations of the flocculent
spiral galaxy NGC~4414 made in order to study the magnetic
field structure in absence of strong density wave flows.
NGC 4414 shows a regular spiral pattern of observed
polarization B-vectors with a radial component comparable in
strength to the azimuthal one. The average pitch angle of the
magnetic field is about 20$\degr$, similar to galaxies with a
well-defined spiral pattern. This provides support for field
generation by a turbulent dynamo without significant
``contamination'' from streaming motions in spiral arms. While
the stellar light is very axisymmetric, the magnetic field
structure shows a clear asymmetry with a stronger regular field
and a smaller magnetic pitch angle in the northern disk.
Extremely strong Faraday rotation is measured in the southern
part of the disk, becoming Faraday thick at 6\,cm. The
distribution of Faraday rotation suggests a mixture of
axisymmetric and higher-mode magnetic fields. The strong
Faraday effects in the southern region suggest a much thicker
magnetoionic disk and a higher content of diffuse ionized gas
than in the northern disk portion. An elongation of the 20\,cm
total power emission is also seen towards the South. Although
NGC 4414 is currently an isolated spiral, the asymmetries in the
polarized radio emission may be sensitive tracers of previous
encounters, including weak interactions which would chiefly
affect the diffuse gas component without generating obvious
long-term perturbations in the optical structure.
 }

\maketitle

\section{Introduction}

The relative role of small-scale velocity perturbations and
galaxy-scale gas flows in determining the evolution and
structure of galactic magnetic fields is one of the most hotly
debated questions in studies of the evolution and
structure of galactic magnetic fields (e.g.
Zweibel~\cite{zwi96}, Zweibel \&~Heiles~\cite{zwi97}, Beck
et al.~\cite{bec96}). Large-scale galactic magnetic fields are
known to display a coherent spiral-like pattern of polarization
B-vectors (see Beck et al.~\cite{bec96} for a review), indicating a
substantial radial component capable of resisting the shear due
to differential rotation. The spiral magnetic field pattern could
be produced by the dynamo mechanism (Wielebinski
\&~Krause~\cite{wie93}, Beck et al.~\cite{bec96}) in which the
small-scale motions are constantly feeding the large-scale
poloidal (i.e. radial and vertical) field. The radial field could also
be produced by continuous field stretching by large-scale flows
due to density waves or bars (e.g. Otmianowska-Mazur
\&~Chiba~\cite{otm95}). In choosing a flocculent spiral, where
flocculent means without large-scale spiral structure, density
waves and bars should play no significant role in
determining the magnetic field orientation.

In well-studied nearby spiral galaxies it is extremely difficult to
discern the effects of the turbulent dynamo from those due to
processes in spiral arms. These objects have well-developed
grand-design patterns with strong density wave compression
and/or a high concentration of star formation in spiral arms. The
first process may strongly modify the magnetic field by effects
of compression and gas flows along the arms (Otmianowska-Mazur
\&~Chiba~\cite{otm95}). In some galaxies like M83
spiral-like compression regions (as traced by aligned dust
filaments) fill the whole interarm space. Strong star formation in
spiral arms acts destructively on regular fields (Beck et
al.~\cite{bec96}). Differences in turbulent activity between the
arms and the interarm region may give rise to strong
concentration of regular magnetic fields between the stellar
arms, thereby strongly influencing the global field structure
(Rohde \&~Elstner~\cite{roh98}).

Knapik et al.~(\cite{kna00}) first reported that flocculent
galaxies may possess regular fields with a strong radial
component. Their study was made at low resolution and only at
one frequency. In this paper we present a high-resolution
multifrequency polarization study of the flocculent galaxy
NGC~4414 (Thornley \&~Mundy~\cite{tho97}). The velocity
field shows no evidence for non-circular gas flows (Braine et
al.~\cite{bra93}, Sakamoto~\cite{sak96}, Thornley
\&~Mundy~\cite{tho97}). NGC~4414 has a high surface gas
density and is forming stars fairly intensively, much like in our
own galaxy, but is in no way a starburst. All this ensures good
conditions for building up a global magnetic field by turbulent
processes in a way unaffected by large-scale gas flows and
compressions.

\begin{table}[th]
 \caption {Basic properties of NGC~4414}
 \begin{flushleft}
 \begin{tabular}{lll} \hline
 Other names & PGC~40692 & Reference\\
 & UGC~7539 & \\
 R.A.$_{1950}$ &$\rm 12^h23^m57\fs 8$& de Vaucouleurs et
al.~(\cite{vau91})\\
 Decl.$_{1950}$ & $31\degr 29\arcmin 58\farcs 0$ &
 de Vaucouleurs et al.~(\cite{vau91})\\
 Inclination & 55$\degr$ & LEDA\\
 Position Angle & 155$\degr$ & LEDA\\
 Distance & 19.2 Mpc & Thim~(\cite{thi00})\\
 Morphol. Type & Sc & LEDA\\
 \hline
\end{tabular} \end{flushleft} \end{table}

\section{Observations and data reduction}

The maps of total power and linearly polarized radio emission of
NGC~4414 at three frequencies were obtained using the Very
Large Array (VLA) of the National Radio Astronomy
Observatory (NRAO)\footnote{NRAO is a facility of National
Science Foundation operated under cooperative agreement by
Associated Universities, Inc.}. At 8.44~GHz and 4.86~GHz the
compact D-array configuration has been used. At 1.415~GHz
and 4.86~GHz observations were made with C-array. The
on-source time was 1.3~hours and 1.7~hours in the C-array
at 1.415~GHz and 4.86~GHz, respectively. In the D-array
the observing time was 14~hours at 8.44~GHz and 13~hours
at 4.86~GHz. The receiver bandwidth was 50~MHz.

The intensity scale was determined by observing 3C286 and
calibrating with the flux densities taken from Baars et
al.~(\cite{baa77}). The position angle of the linearly polarized
emission was calibrated using the same source with an assumed
position angle of 33$\degr$. The phase calibrator 1219+285 was
also used to determine telescope phases and the instrumental
polarization.

The data reduction has been performed using the AIPS data
reduction package. The edited and calibrated visibility data were
Fourier transformed to obtain maps of the Stokes parameters I,
Q and U at all three frequencies. At 4.86~GHz the visibilities
from both array configurations were combined. The maps were
weighted with ROBUST=0 for the best resolution and sensitivity
compromise, yielding synthesized half-power beam widths
(HPBW) of 7$\arcsec$ and 11$\arcsec$ at 8.44~GHz and
4.86~GHz, respectively. At 1.415~GHz our C-array map also
weighted with ROBUST=0, has an original HPBW of
14$\arcsec$. The naturally weighted maps, more sensitive to
extended structures, have HPBWs of 11$\arcsec$ and
16$\arcsec$ at 8.44~GHz and 4.86~GHz, respectively. The Q
and U maps were combined to get maps of the linearly polarized
emission (corrected for the positive zero level offset) and of the
position angle of apparent magnetic vectors (B-vectors).

NGC~4414 has been also observed using the Effelsberg 100-m
MPIfR telescope to determine the total power flux at
2.695~GHz. Several small maps of the galaxy scanned either in
R.A. or in declination were made, then the maps were averaged
and the flux density was determined by integrating the final map
in concentric rings.

\section{Results}

\subsection{Total power emission}
%================================
\begin{figure}
\resizebox{\hsize}{!}{\includegraphics{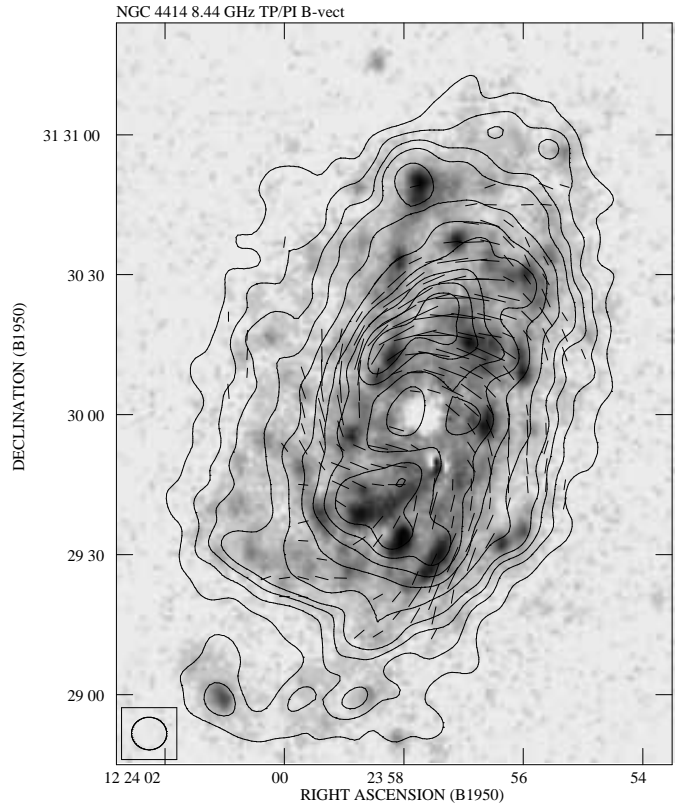}}
\caption{
Total power map of NGC~4414 at 8.44~GHz with polarization
B-vectors superimposed onto the H$\alpha$ image from
Pogge~(\cite{pog89} and priv. comm.). Uniform weighting
yielding a resolution of 7$\arcsec$ and r.m.s. noise of
9~$\mu$Jy/b.a. has been applied. Contour levels are 30, 70, 110,
150, 250, 400, 550,\ldots $\mu$Jy/b.a. }
\label{fig01}
\end{figure}
%================================
\begin{figure}
\resizebox{\hsize}{!}{\includegraphics{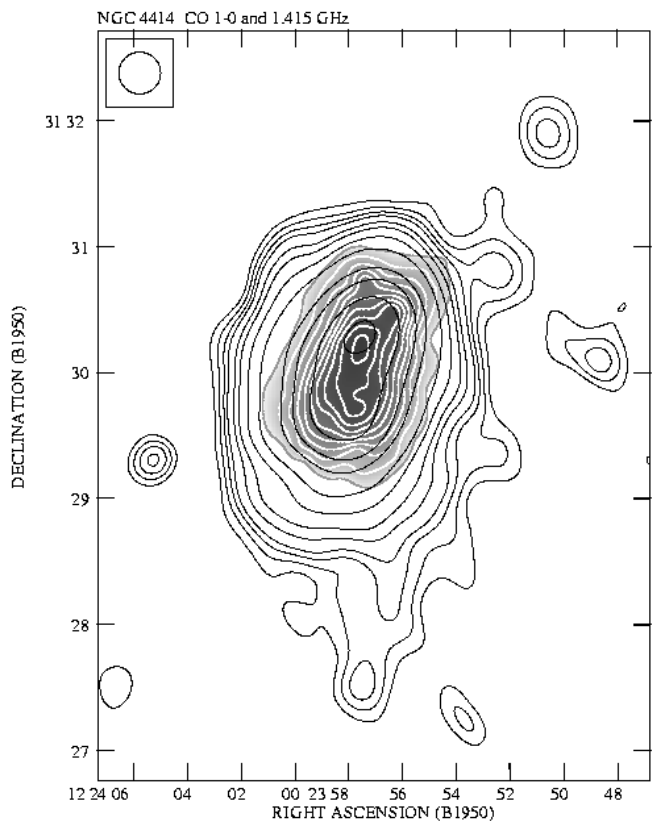}}
\caption{
The contour map of total power emission from NGC~4414 at
1.415~GHz. The contour levels are 2, 3, 4, 5, 7, 9, 13, 20, 40,
60, 100, 150, 200 $\times$ 170$\mu$Jy/b.a. The map has been
smoothed to a resolution of $20\arcsec$. The map is overlaid
upon the distribution of CO(1--0) emission (Braine et
al.~\cite{bra93}) shown as white contours on a grayscale
background
}
\label{tp20}
\end{figure}

%==================================

\begin{figure}
\resizebox{\hsize}{!}{\includegraphics{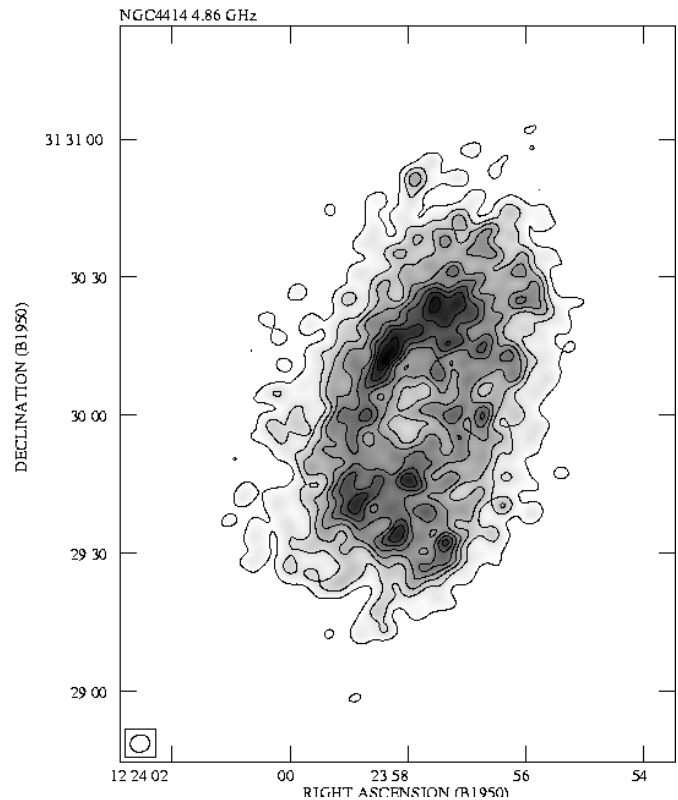}}
\caption{
The total power map of NGC~4414 at 4.86~GHz with a
resolution of $4\arcsec$ with a greyplot of the same quantity.
The contour levels are 3, 5, 7, 9, 11, 13, 15, 17, 19, 21, 23, 25,
27 $\times$ 7~$\mu$Jy/b.a. }
\label{hires}
\end{figure}
%========????????=================

The contour map of the total power brightness of NGC~4414 at
8.44~GHz is presented in Fig.~\ref{fig01}. The total power
emission shows a ring-like distribution with a central depression
roughly at the position of the hole in the CO and H$\alpha$
emission (Thornley \&~Mundy~\cite{tho97},
Sakamoto~\cite{sak96}, Braine et al.~\cite{bra93}). Our total
power map also shows two maxima. The southern maximum
coincides roughly with an optically bright clump and with a
group of three large \ion{H}{ii} regions (Pogge~\cite{pog89}).
The peanut-shaped northern peak extends eastwards from the
major axis and forms a bridge between two groups of
\ion{H}{ii} regions. These brightness maxima have almost the
same amplitude relative to the rest of the disk in our total power
maps at 1.415~GHz and at 8.44 GHz. They are thus certainly
nonthermal. In the outermost disk the radio emission in our total
power maps at 8.44 and 4.86 GHz is asymmetric with a steeper
brightness gradient along the western disk boundary
(Fig.~\ref{fig01}). A small peak north of the centre coincides
with a bright star-forming region and is not visible in our
1.415~GHz map so we suspect that it has an increased fraction
of thermal emission. A faint radio ridge extends eastwards from
the southern end of the major axis. It coincides with weak
optical and H$\alpha$ extensions. At low frequencies, the total
power map at 1.415~GHz also shows a clear tail extending to
the south (Fig.~\ref{tp20}).

Our C-array map of NGC~4414 at 4.86~GHz shows the
brightest total power peaks with a resolution of $4\arcsec$
(Fig.~\ref{hires}). The total power hole in the disk centre is very
deep at this resolution with no trace of any nuclear source. In the
southern disk there is a good agreement between the total power
peaks and bright \ion{H}{ii} regions. In the outer parts of the
NE disk a large \ion{H}{ii} region at R.A.$_{1950}=\rm 12^h
23^m 57\fs 9$ Dec$_{1950}=+31\degr 30\arcmin 50\arcsec$
corresponds to a local radio emission maximum as well.
However, in the bright peanut-shaped feature the association is
less obvious: its radio-brightest part lies between nearby
H$\alpha$ peaks.

%================================
\begin{figure}
\resizebox{\hsize}{!}{\includegraphics{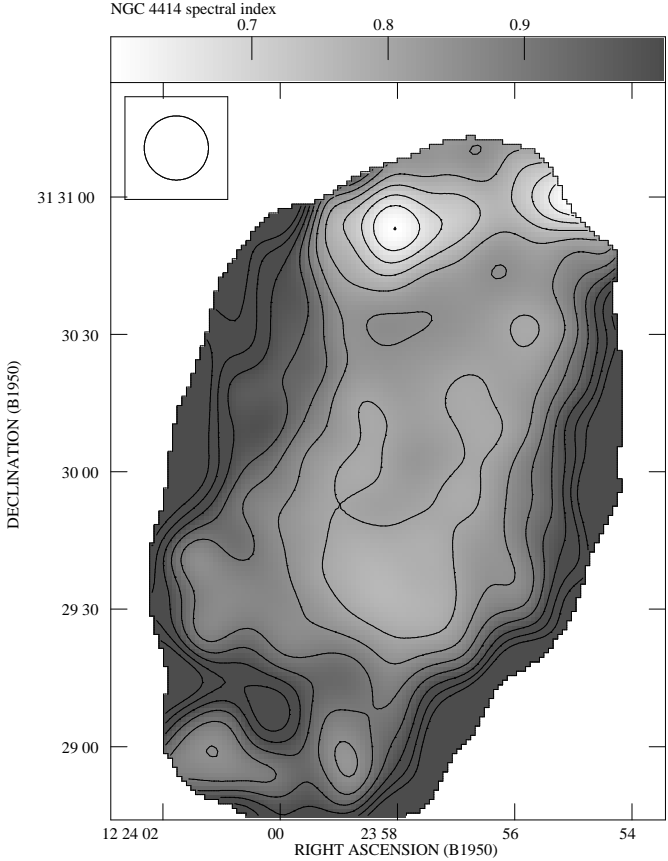}}

\caption{
The spectral index distribution over the disk of NGC~4414 fitted
to all three frequencies. The contour levels start from 0.55 with a
step of 0.05. All the maps are convolved to a beam of 14\arcsec
}
\label{spix}
\end{figure}
%==================================

The spectral index distribution in the disk of NGC~4414 is
shown in Fig.~\ref{spix}. The flattening of the spectrum in the
NE disk corresponds to a bright complex of \ion{H}{ii} regions
at R.A.$_{1950}=\rm 12^h 23^m 57\fs 9$
Dec$_{1950}=+31\degr 30\arcmin 50\arcsec$. The mean
spectral index in the inner disk (within the radius of $1\arcmin$)
is about $\alpha \approx 0.82$ ($S_{\nu}\propto \nu^{-\alpha}$)
to the North $\alpha \approx 0.79$ to the South. The difference
is insignificant (r.m.s variations are about 0.02) and may be due
to the flattening to $\alpha \approx 0.76$ at the position of large
\ion{H}{ii} regions in the southern disk. Thus, there is no clear
evidence for a flatter spectrum of total power emission in the
southern disk region.

\subsection{High-frequency polarized intensity}

%================================
\begin{figure}
\resizebox{\hsize}{!}{\includegraphics{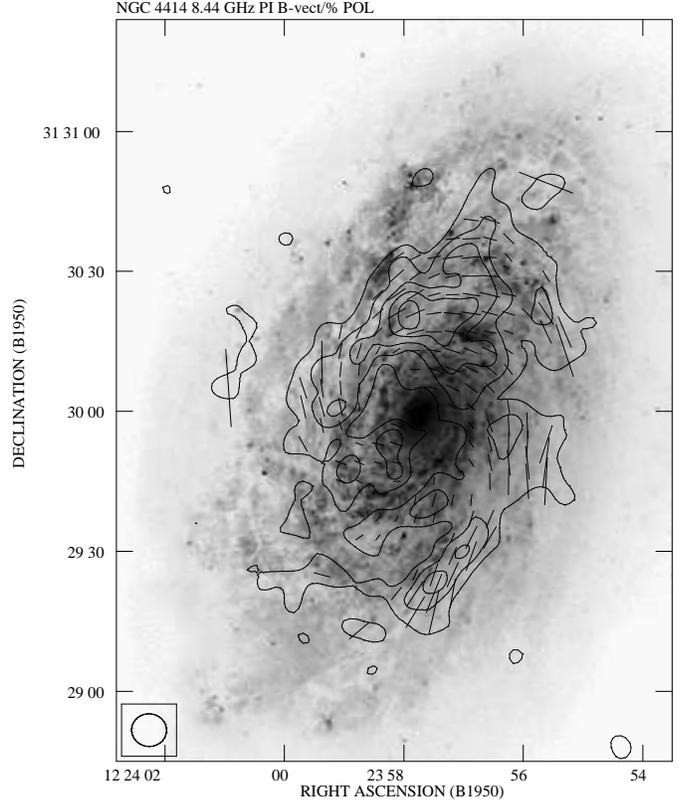}}
\caption{
Contour map of the polarized intensity of NGC~4414 at
8.44~GHz with
a resolution of 7$\arcsec$ (uniform weighting for highest
resolution) and
r.m.s. noise of 8~$\mu$Jy/b.a. with B-vectors proportional to the
polarization degree, overlaid upon the optical image from HST
(permission from Dr W. Freedman from Carnegie Institution).
Contours
are drawn every 25~$\mu$Jy/b.a.
}
\label{fig02}
\end{figure}
%================================

The high resolution map of the polarized intensity at 8.44~GHz
is presented in Fig.~\ref{fig02}. The brightest peak of polarized
emission is located in the northern half of the disk, east of the
bright ensemble of \ion{H}{ii} regions. An elongated polarized
feature is also located in the southern disk, just outside of the
optically bright arm-like patch. The peaks of high polarized
intensity do not clearly avoid nor correspond to bright star-forming
regions. Diffuse, more smoothly distributed polarized
emission is visible as well.

\begin{figure}
 \resizebox{\hsize}{!}{\includegraphics{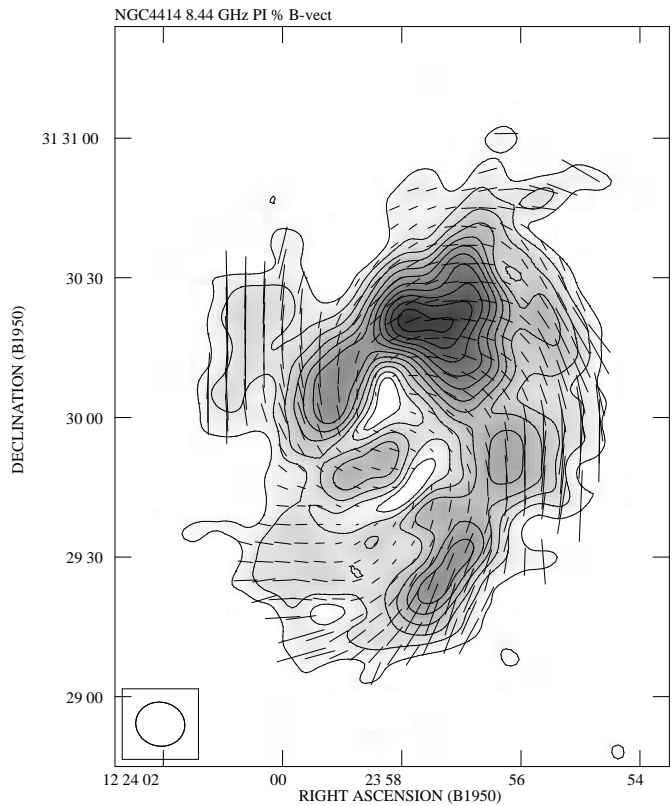}}
 \caption{
Contour map of the polarized intensity of NGC~4414 at
8.44~GHz with B-vectors proportional to the polarization degree
of NGC~4414. Natural weighting yielding an r.m.s. noise of
7~$\mu$Jy/b.a., a beam of 11$\arcsec$ and a higher sensitivity
to extended structures has been applied. Contours are plotted
every 25~$\mu$Jy/b.a. The vector of $10\arcsec$ length
corresponds to the polarization degree of 50\%
}
\label{fig03}
\end{figure}

Despite the lack of organized optical or H$\alpha$ spiral
structure the polarization B-vectors form a clear spiral pattern.
Our 8.44~GHz polarization map at high resolution
(Fig.~\ref{fig02}) shows that the magnetic field orientation does
not fluctuate from place to place, being coherent even at the
scale of our smallest beam (650~pc). The global magnetic spiral
pattern becomes fully visible when natural weighting is applied,
lowering the resolution to 11$\arcsec$ but allowing smooth,
low-brightness regions to be seen (Fig.~\ref{fig03}). The B-vectors
are nearly azimuthal (pitch angle near 0$\degr$) close to
the northern semi-axis but show a strong radial component
(pitch angle up to 45$\degr$) in the southern disk. In the very
inner disk the polarization B-vectors cross the galaxy's centre
along the minor axis, forming an S-shaped pattern. On each side
of the centre polarization minima are visible, presumably caused
by beam depolarization of strongly twisted magnetic fields, seen
perpendicular in sky projection.

The highest degrees of polarization (up to some 30\% at
8.44~GHz) were found along the western disk boundary and in
the faint eastern tail. The inner disk shows a significant north-south
asymmetry. The mean polarized brightness at 8.44~GHz
and the mean polarization fraction are about 40\% higher in the
North than the South, where the H$\alpha$ emission is slightly
stronger. In particular, the H$\alpha$ map (Pogge~\cite{pog89}
and priv. comm.) convolved to the beam of 23\arcsec \, (the
resolution of the CO(1--0) map) shows the brightest peak in the
southern disk. The mean H$\alpha$ brightness, within the area
delineated by the level of 30\% of the peak value, is some 10\%
higher in the southern disk than in the northern one. In this
respect NGC~4414 resembles nearby galaxies in which the
highly polarized emission tends to avoid regions of high star
formation (Beck et al.~\cite{bec96}).

%==================================
%==================================

\subsection{Faraday rotation and depolarization}

%=================================
\begin{figure}
 \resizebox{\hsize}{!}{\includegraphics{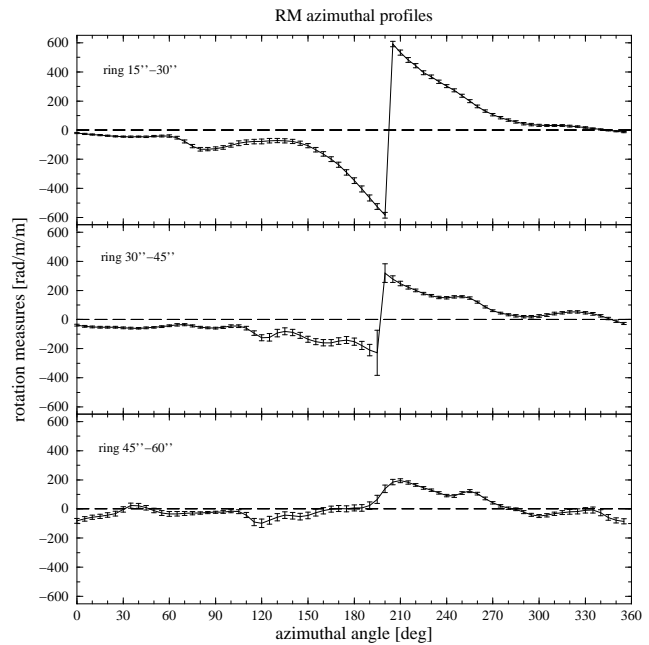}}
 \caption{
Azimuthal profiles of the Faraday rotation measures of
NGC~4414 between 4.86~GHz and 8.44~GHz integrated in
sectors of azimuthal width of $5\degr$ along rings 15$\arcsec$
wide with the same inclination and position angle as the galaxy.
The azimuthal angle runs counterclockwise from the northern
end of major axis. The naturally weighted maps were convolved
to a common beam of 16$\arcsec$. Angles have been corrected
to face-on}
\label{fig04}
\end{figure}
%====================================

NGC~4414 lies very close to the North Galactic Pole where the
foreground Faraday rotation measure (RM) is smaller than
30~rad/m$^2$ (Simard-Normandin
\&~Kronberg~\cite{sim80}). Using the polarized background
source found in our maps at about $4\arcmin$ NE from
NGC~4414 we determined the foreground rotation to be
$-4\pm8$~rad/m$^2$. For this reason no correction for the
foreground rotation was applied to our data.

The changes of RM with azimuthal angle in the disk
(Fig.~\ref{fig04}) do not show clearly any simple singly-periodic
or doubly-periodic variations traditionally attributed to
axisymmetric (ASS) or bisymmetric (BSS) global field
symmetry (Beck et al.~\cite{bec96}). Instead, the azimuthal
profiles show a sudden jump at an azimuthal angle of about
200$\degr$ with RM values in the inner disk changing abruptly
from $+600$~rad/m$^2$ to $-600$~rad/m$^2$. The rapid sign
change occurs almost throughout the southern half of the disk.
Such a jump usually means that the Faraday rotation angle
exceeds 90$\degr$, corresponding at 4.86~GHz to RM$ =
\pi/2\lambda^2$ $\simeq 600$~rad/m$^2$. It does not imply
any reversal of the magnetic field in NGC~4414 but signifies a
``Faraday-thick regime'' when the rotation angle exceeds
90$\degr$ (Sokoloff et al.~\cite{sok98}). Such strong Faraday
rotation is exceptional at 4.86~GHz and has not been observed
in moderately inclined nearby galaxies which become ``Faraday-thick''
much below this frequency (Beck at al.~\cite{bec96}).
The jump is most conspicuous at small galactocentric radii,
becoming smoother and of smaller amplitude in the outer disk.
The azimuthal profiles shown in Fig.~\ref{fig04} also
show that RM values in the northern half of the disk are of
order of 50--70 rad/m$^{2}$, much smaller than in the southern
half, especially at the galactocentric radii $<45\arcsec$.

\begin{figure}
 \resizebox{\hsize}{!}{\includegraphics{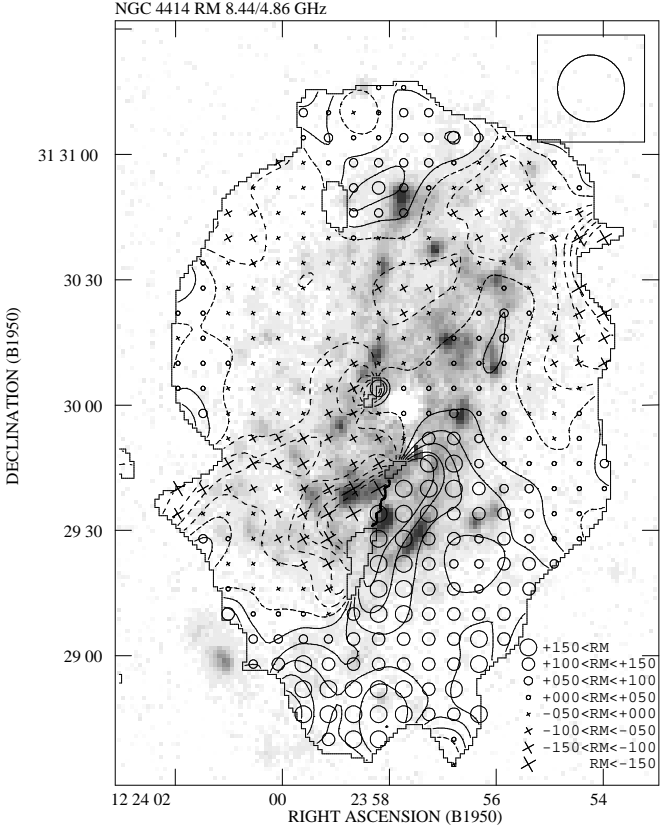}}
 \caption{
The distribution of rotation measures of NGC~4414 between
4.86~GHz and 8.44~GHz, overlaid on the greyscale H$\alpha$
image (Pogge~\cite{pog89} and priv. comm.) showing the
brightest \ion{H}{ii} regions. Circles denote positive and
crosses negative RM. The symbol sizes depend on the absolute
value of RM as shown in the legend. The contours show
negative (dashed) and positive (solid) values of RM with a step
of 50~rad$/$m$^2$. The naturally weighted maps at both
frequencies were convolved to a common beam of 16$\arcsec$
}
\label{fig05}
\end{figure}

Details of the distribution of Faraday rotation between
8.44~GHz and 4.86~GHz are shown in Fig.~\ref{fig05}. The
rotation measure (RM) forms large intermixed domains of
positive and negative values which do not correspond to any of
the known global field symmetries. Generally, positive values
dominate in the SW and western disk part with a small spot at
the northern tip of the major axis. In large parts of the southern
and SW disk the RM exceeds $+100$~rad/m$^2$. A narrow
region of very strong Faraday rotation, reaching $+600$
rad/m$^{2}$ is located in the southern disk close the
depolarized region south of the centre, jumping on its other side
down to $-600$~rad/m$^2$. The negative rotation measures
occupy the SE and eastern disk regions, extending to northwest.
In the northern disk the rotation measures are small ($\leq
100$~rad/m$^2$) with interspersed domains of positive and
negative sign. This region is also substantially polarized at
1.415~GHz, yielding similar values of RM determined between
this frequency and 8.44~GHz. The above picture only weakly
depends on the assumption concerning small
($<30$~rad/m$^2$) foreground Faraday rotation.

%================================
%================================

\begin{figure}
\resizebox{\hsize}{!}{\includegraphics{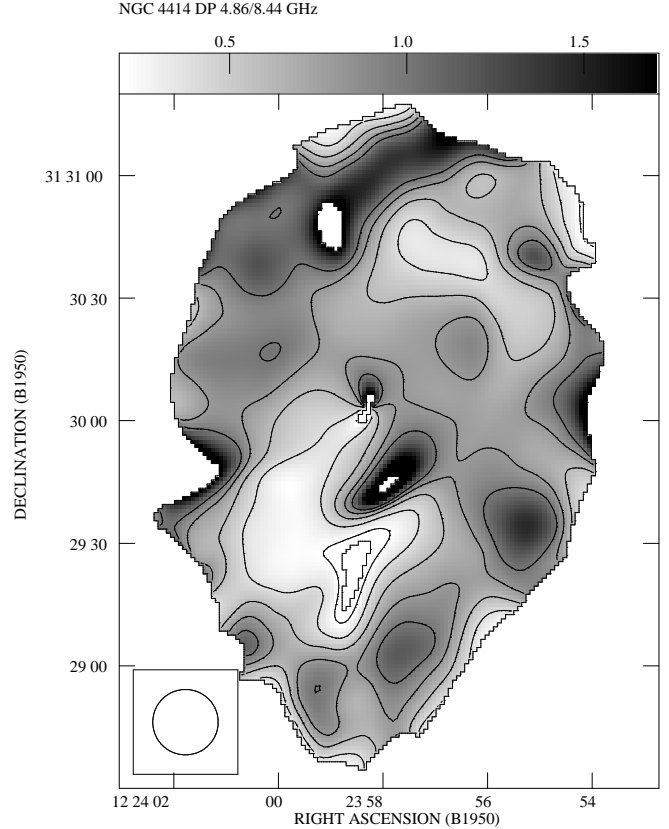}}
\caption{
The distribution of Faraday depolarization (DP) in NGC~4414
between 4.86~GHz and 8.44~GHz determined from naturally
weighted maps convolved to 16$\arcsec$. The contour levels are
0.2, 0.4, 0.6, 0.8, 1.0
}
\label{fig06}
\end{figure}

The map of Faraday depolarization (DP) between 8.44 and
4.86~GHz (defined as the ratio of polarization degree between
the lower and higher frequencies) is shown in Fig.~\ref{fig06}.
In the southern disk a region strongly depolarized at 4.86~GHz
is present, with a depolarization factor DP $\le$ 0.2. It is
associated with a jump in RM, as expected for a depolarization
due to strong Faraday rotation inside the emitting region (see
Burn~\cite{bur66}, Sokoloff et al.~\cite{sok98}) and extends to
the SE disk region. The southern disk is completely depolarized
at 1.415~GHz which confirms the strong Faraday effects.
Another moderately depolarized region was found NW of the
centre, close to the major axis. The DP there is about 0.45, rising
to 0.5--0.6 when a correction for depolarization due to RM
gradients across the beam is applied. On average, the DP
between 4.86 and 8.44~GHz in northern disk is about 0.7. This
region was even detected in polarization at 1.415~GHz thus, it is
definitely Faraday-thin at 4.86~GHz.

The regions of strong Faraday rotation or depolarization show
some association with the ionized gas, as traced by the
H$\alpha$ line. Both the Faraday rotation jump and the
depolarized area lie close to the complex of three large
\ion{H}{ii} regions and the jump south of the centre coincides
with the strongest peak of H$\alpha$ emission. The area
depolarized by more than 50\% close to the northern major
semi-axis lies close to two large \ion{H}{ii} regions. However,
a similar complex of ionized gas NW of the disk centre (around
R.A.$_{1950}=\rm 12^h 23^m 56\fs 3$
Dec$_{1950}=+31\degr 30\arcmin 13\arcsec$) shows a
rather low RM and is not strongly depolarized at
4.86~GHz.

%================================
%==================================

\section{Discussion}

\subsection{Total magnetic field strength}

\begin{table}[th]
 \caption {The integrated radio spectrum of NGC~4414}
 \begin{flushleft}
 \begin{tabular}{rrrl} \hline
 Frequ-&Flux & error & References \\
 ency &density& & \\\
 [GHz] & [mJy] & [mJy] & \\ \hline
 0.408 & 560 & $\pm$ 25& Gioia \&~Gregorini~(\cite{gio80})\\
 1.490 & 231 & $\pm$ \,\ 2& Condon at al.~(\cite{con90})\\
 2.695 & 134 & $\pm$ 30& This work.\\
 4.800 & \ 76 & $\pm$ \,\ 5& Gioia at al.~(\cite{gio82})\\
 4.850 & \ 78 & $\pm$ \,\ 5& Condon at al.~(\cite{con91})\\
 4.860 & \ 83 & $\pm$ \,\ 5& This work.\\
 10.55 & \ 58 & $\pm$ \,\ 5& Niklas at al.~(\cite{nik95})\\
 10.70 & \ 43 & $\pm$ \,\ 7& Gioia at al.~(\cite{gio82})\\
 \hline
\end{tabular}
\end{flushleft}
\end{table}

The flux densities of NGC~4414, compiled from data available
in the literature are shown in Tab. 2. The value at 2695~MHz
has been obtained by integrating our Effelsberg map in circular
rings (the galaxy was only barely resolved) out to a radius of
$6\arcmin$. The flux density at 4.86~GHz was obtained from
the integration of our VLA naturally weighted total power map
in elliptical rings with the inclination and position angle from
Tab. 1 out to a radius of $3\arcmin$.

The best power-law fit to this data yields $\alpha $= 0.77. Using
this value and assuming energy equipartition, a lower limit to the
cosmic-ray spectrum to be 300~MeV (see Beck~\cite{bec91}), a
ratio of proton-to-electron energy density of 100
(Pacholczyk~\cite{pah70}), and a nonthermal disk thickness of
2~kpc (full thickness corrected to face-on), we compute the
mean magnetic field strength within the radius delineated by
5\% of the maximum brightness at 4.86~GHz to be
$<B_t>=15\pm 4~\mu$G. The uncertainty takes into account
factor two variations in the proton-to-electron ratio, the disk
thickness and the lower energy cutoff as well as, a variation in
the thermal fraction at 4.86~GHz between 0 and 10\% (allowed
by spectral fits to the data in Tab. 2). The total magnetic field in
NGC~4414 is comparable to that of actively star-forming (non-starburst)
spiral galaxies (Beck et al.~\cite{bec96}). Assuming
the regular field to be parallel to the disk plane we obtain its
mean strength $<B_{reg}>=4\pm 1\mu$G. The average ratio
$<B_{reg}>/<B_t>$ is 0.25.

\subsection{Distribution of the nonthermal emission}

%=============================
\begin{figure}
\resizebox{\hsize}{!}{\includegraphics{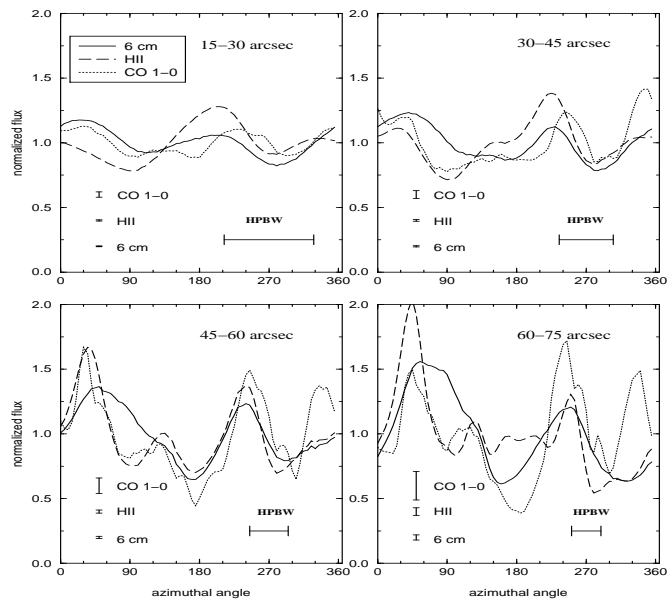}}
\caption{
Azimuthal variations of the total power brightness of
NGC~4414 at 4.86~GHz, compared to those of CO(1--0)
(Braine et al.~\cite{bra93}) and H$\alpha$
(Pogge~\cite{pog89}) averaged in sectors with $5\degr$
azimuthal width along concentric rings 15$\arcsec$ wide. All
the data were convolved to a common beam of 23$\arcsec$. The
profiles were normalized to the mean value in each ring. The
disk was rectified to face-on. Horizontal bars denote the beam
size
}
\label{fig07}
\end{figure}
%============================

%=============================
\begin{figure}
\resizebox{\hsize}{!}{\includegraphics{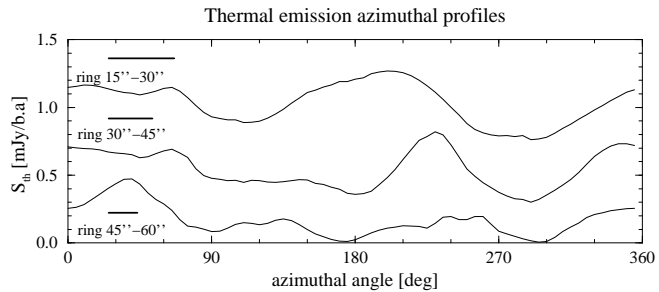}}
\caption{
Azimuthal variations of the thermal emission of NGC~4414 at
8.44~GHz, averaged in sectors with $5\degr$ azimuthal width
along concentric rings 15$\arcsec$ wide. All the data were
convolved to a common beam of 16$\arcsec$. A nonthermal
spectral index of 1 has been assumed. The disk was rectified to
face-on. Thick horizontal bars denote the beam size
}
\label{sth}
\end{figure}
%============================

Significant correlations between the gas density, star
formation rate and magnetic fields were established
by Berkhuijsen et al.~(\cite{ber93}). In grand-design
objects these quantities are affected by processes in spiral
arms, like the accumulation of molecular gas in dust lanes
followed by massive star formation and the magnetic field
enhancement in compression regions.

Azimuthal distributions of the 4.86~GHz total power brightness
of NGC~4414 (showing a smaller contamination by thermal
emission than that at 8.44~GHz and less affected by the zero-spacing
problem than the C-array map at 1.415~GHz) are
compared to the H$\alpha$ and CO(1--0) emission in
Fig.~\ref{fig07}. At all radii there is a reasonable
correspondence between the maxima of emission at 4.86~GHz
and those in the H$\alpha$ and CO lines. In the innermost disk
region the correlation is less obvious and the maxima shifted,
possibly because of severe beam-smearing effects. At larger
radii the azimuthal variations of the total power brightness
follow well those in CO and in H$\alpha$. The maximum of the
total power brightness at the azimuthal angle of about $45\degr$
is visible both in CO and H$\alpha$. However, instead of a
second maximum visible in CO and H$\alpha$ at the azimuth of
$\simeq 120\degr$ the radio emission shows only an asymmetric
extension. It is possible that the CO and H$\alpha$ minimum at
the azimuth of $90\degr$ has been smoothed out by cosmic-ray
propagation effects. All profiles exhibit a similar rise for
azimuthal angles $> 180\degr$. However, the CO peak at an
azimuthal angle of about 340$\degr$ in the rings outside a radius
of $30\arcsec$ does not correspond to a similar feature in the
H$\alpha$ line. It has no counterpart in the radio domain so the
H$\alpha$ emission is not caused by strong absorption of the
optical radiation from hidden star-forming regions. Thus,
the interrelations between gas density, star formation rate
and magnetic fields also exist in flocculent galaxies, in which
the distributions of gas, young stars and magnetic field are
not strongly modulated by density waves. They have a
global character, while small-scale details may show
substantial place-to-place differences.

%==============================
\begin{figure*}[htpb]
\resizebox{12cm}{!}{\includegraphics{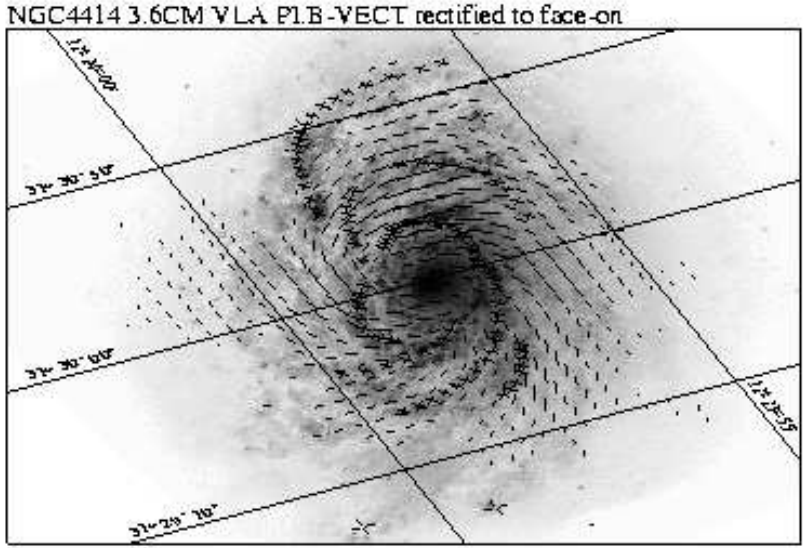}}
\parbox[b]{55mm}{
\caption{
B-vectors of polarized intensity of NGC~4414 overlaid upon the
optical image from HST (permission from Dr W. Freedman
from Carnegie Institution), both rectified to face-on view. The
optical image was digitally filtered to enhance the spiral pattern.
The spiral armlets discussed in Thornley and
Mundy~(\cite{tho97}) are marked by symbols
}
\label{fig10}}
\end{figure*}
%===============================

The central depression in the radio emission, gas density and star
formation deserves attention. The gas in this region is depleted
during the early strong star formation necessary for the high
stellar densities found in galaxy nuclei. In density wave galaxies
the gas is replenished by radial flows in spiral arms, which feed
star-forming processes in the centre and inner disk. Some
flocculent galaxies show an accumulation of CO gas in the
central region (Regan et al.~\cite{reg01}) but this is not the
case for NGC~4414 (Sakamoto~\cite{sak96}). The central
minimum of all species in NGC~4414 is an argument for the
absence of dynamically important radial gas flows.

\subsection{Distribution of the thermal emission}

%===============================

\begin{figure*}[htpb]
\resizebox{12cm}{!}{\includegraphics{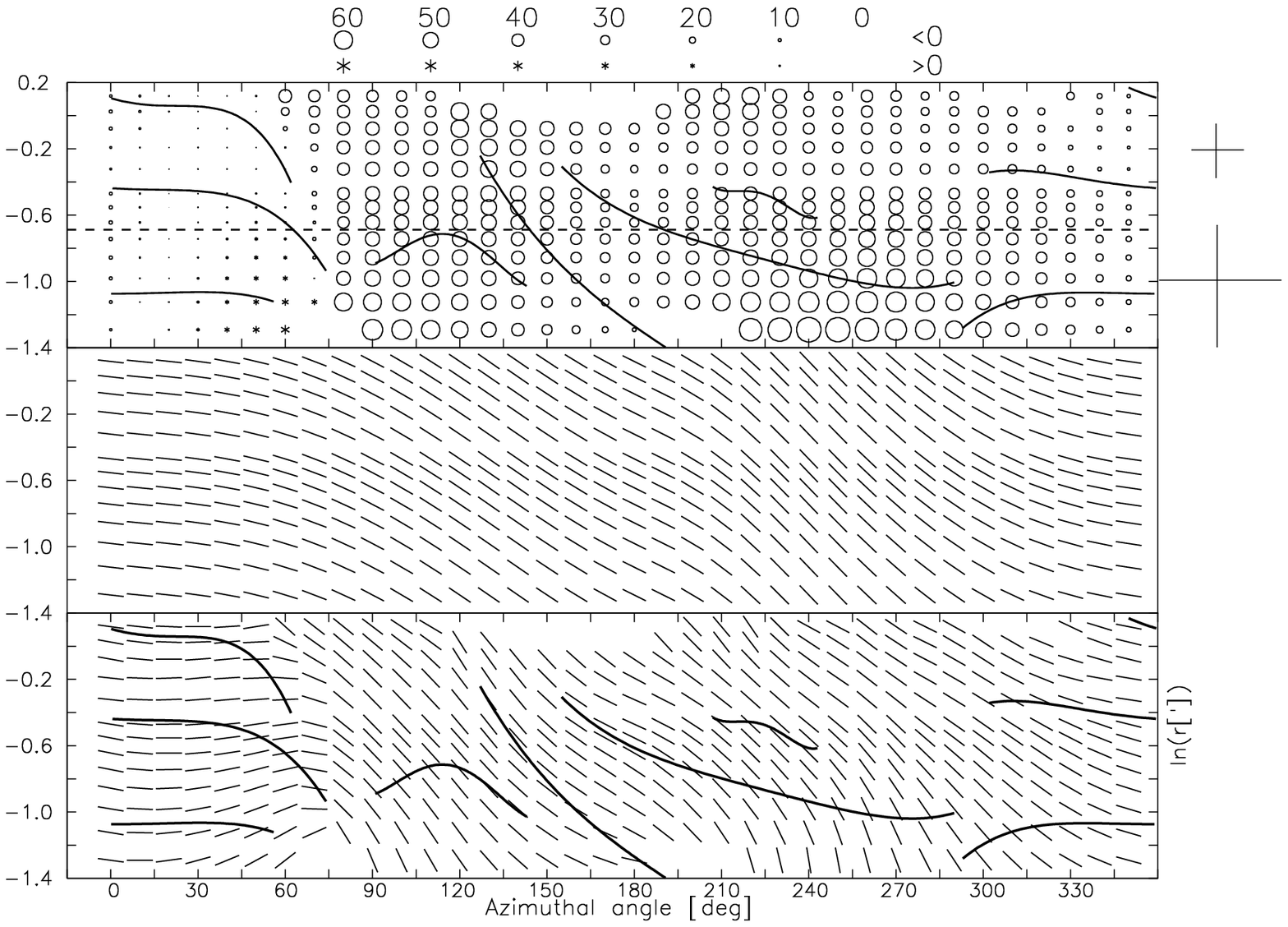}}
\parbox[b]{55mm}{
\caption{
Top: the distribution of absolute values of pitch angles as a
function of azimuthal angle and $\ln(r)$, $r$ being the
galactocentric distance in arcmin. The azimuthal angle runs
counterclockwise from the northern major semi-axis. Circles
denote negative and stars positive values. The dashed line
marks the radius beyond which NGC~4414 rotates
differentially (Sakamoto~\cite{sak96}). Middle: the
orientations of magnetic B-vectors for a mixture of m=0, m=1
and m=2 modes as described in the text, compared to the
observed orientations of magnetic B-vectors in NGC~4414 at
8.44~GHz (bottom panel). Crosses on the right side of the graph
show the beam size for two different distances from the galaxy's
centre. Solid lines show the spiral armlets discussed by Thornley
\& Mundy~(\cite{tho97})
}
\label{fig11}}
\end{figure*}
%=============================

The interpretation of our Faraday rotation data needs
information on the distribution of thermal electron density
independent of the H$\alpha$ data which suffer from extinction
effects. The azimuthal distribution of thermal emission from
NGC~4414 at 8.44~GHz (Fig.~\ref{sth}) was determined from
our spectral index map (Fig.~\ref{spix}) assuming the
nonthermal spectral slope $\alpha_{nt}$ constant over the whole
disk and equal to 1.0 -- the mean observed value of $\alpha$ in
the disk outskirts. The detailed value of $\alpha_{nt}$ has little
meaning for our discussion as it can only shift vertically the
curves shown in Fig.~\ref{sth} without changing the shape of
azimuthal variations of the thermal flux.

The distribution of the thermal emission shows two maxima of
similar amplitude at both ends of the major axis. The northern
maximum at an azimuthal angle of about $30\degr$ coincides
well with a similar peak of total radio power, CO and H$\alpha$.
The southern maximum is only slightly stronger and roughly
corresponds to a second peak of all these species at an azimuth
$\simeq 220\degr$--$240\degr$. We do not find any strong
excess of thermal emission in the southern disk which would
naturally explain why the Faraday effects are so much stronger
than in the northern disk.

The lack of much higher thermal emission in the southern disk
also means that this region does not exhibit strong star formation
hidden by absorption in the H$\alpha$ line. Thus we cannot
explain the lower degree of polarization at high-frequency of the
southern region by stronger magnetic field tangling generated by
star-formation. The depolarization by differential Faraday
rotation at 8.44~GHz reaches 0.68 at the RM jump. However,
such strong depolarization occupies a small fraction of the disk
and cannot account for the systematically lower mean
polarization degree of the southern disk.

\subsection{The magnetic field structure}

The CO velocity field observed by Sakamoto~(\cite{sak96})
does not show any perturbations similar to density wave
streamings (Visser~\cite{vis80}), which would cause radial
stretching of magnetic field lines. As also stated by Thornley
\&~Mundy~(\cite{tho97}) the optical filaments or armlets
which they detected in the near infrared do not cause strong
departures from otherwise smooth HI and CO velocity fields.
Though the armlets are likely to be dynamical features, they
do not constitute a well-organized grand-design spiral
pattern (Thornley \&~Mundy~\cite{tho97}). Small-scale
phenomena like chaotic, turbulent gas motions may play an
important role in the dynamics and star-forming processes
in NGC~4414.

Despite the lack of organized spiral arms, the B-vectors
(rectified to face-on) (Fig.~\ref{fig10}) form a very regular
spiral pattern. The magnetic pitch angle $\psi$ observed in
NGC~4414 is about zero in the northern disk, increasing to
about $40\degr$--$45\degr$ over the rest of the disk. This is also
visible in the azimuth-$\ln(r)$ frame (Fig.~\ref{fig11}) in which
a logarithmic spiral would form a straight line inclined by the
pitch angle. The infrared armlets detected by Thornley \&
Mundy (\cite{tho97}) show only limited similarity to the
magnetic field structure. They generally have smaller pitch
angles in the northern disk and larger ones in the remaining disk
portions. However, the magnetic field is aligned only with some
portions of armlets while the polarization B-vectors apparently
run across other armlet segments in all parts of the disk
(Fig.~\ref{fig10}, Fig.~\ref{fig11}). For the deprojection to
face-on, we assumed that both the armlets and magnetic vectors
lie in the galaxy plane. As demonstrated by Knapik et
al.~(\cite{kna00}) the presence of a three-dimensional magnetic
field geometry as generated by the dynamo does not strongly
affect the orientation of deprojected B-vectors even at
inclinations as high as $60\degr$. Thus, while the armlets may
still have some influence upon the magnetic field structure, local
processes like field amplification by the turbulent dynamo
process need to be considered.

\begin{figure}
\resizebox{\hsize}{!}{\includegraphics{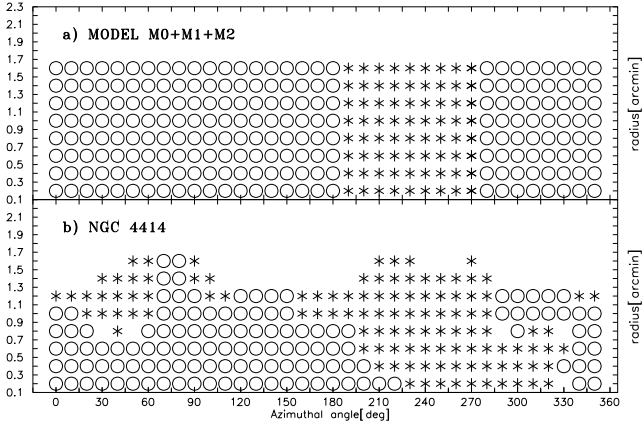}}
\caption{
Distribution of the sign of the line-of-sight magnetic field
(responsible for Faraday effects) in our model (a) assuming a
mixture of m=0, m=1 and m=2 modes compared to (b) the sign
of Faraday rotation between 8.44~GHz and 4.86~GHz for
NGC~4414. The Q and U data were convolved to a beam of
$21\arcsec$ and restricted to signal stronger than $2\sigma$
r.m.s.
}
\label{rmmodl}
\end{figure}

The magnetic vectors crossing the disk centre are very
suggestive of an admixture of non-axisymmetric magnetic field
components, especially as the field running through the centre
along the minor axis cannot be explained by the limited angular
resolution (see Knapik et al.~\cite{kna00}). The hypothesis of a
mixture of modes is supported by an analysis which
simultaneously uses position angles of B-vectors at 8.44~GHz
and 4.86~GHz (thus involving Faraday rotation effects), made
kindly for us by Dr Andrew Fletcher (priv. comm). The
azimuthal variations of magnetic pitch angles are best
reproduced by a mixture of axisymmetric (the azimuthal
magnetic wavenumber m=0), bisymmetric (m=1) and m=2 fields
(Fig.~\ref{fig11} middle panel). Because the magnetic field is a
vector (in contrast to e.g. density), for the BSS configuration
m=1 is required to get opposite signs of magnetic field
components at azimuthal angles differing by $180\degr$. The
amplitudes of the higher modes relative to the axisymmetric
field are 0.6 (m=1) and about 0.3--0.5 (m=2, varying with the
radius). The axisymmetric field used in this analysis forms a
trailing spiral with a pitch angle $\psi$ of about $22\degr$, the
bisymmetric one has $\psi$ of only $2\degr$ while the m=2
component has a large pitch angle whose absolute value varies
between $40\degr$ and $80\degr$. This mixture of modes well
explains the highest polarization extending from the northern
major semi-axis along the western disk edge (see
Fig.~\ref{fig03}) and the weakest polarized intensity in the SE
disk. It also yields two asymmetric regions of a constant sign of
magnetic field along the line of sight (responsible for Faraday
rotation) with the sign reversal at the azimuthal angles of about
$180\degr$ and $270\degr$. This bears some resemblance to the
RM sign distribution in NGC~4414 at radii $<1\arcmin$
(Fig.~\ref{rmmodl}).

%=============================

%=============================

As suggested by Chiba~(\cite{chi93}) and
Moss~(\cite{mos96a}) the generation of bisymmetric magnetic
fields may be boosted by the parametric resonance in strong
density waves. In this work we show the evidence for higher
modes of the field structure {\it without strong effects of spiral
arms}. The apparent contribution of an m=2 mode may indicate
a modulation of the lower modes e.g. due to the non-axisymmetric
distribution of thermal gas (see Fig.~\ref{sth})
and Faraday effects. Furthermore, the models discussed assume
only a two-dimensional magnetic field parallel to the disk. Their
further development requires a three-dimensional divergence-free
magnetic field, computed using elaborate MHD models.

Though NGC~4414 is currently an isolated spiral (e.g. Braine et
al.~\cite{bra97}), the observed magnetic field asymmetries may
also signify external interactions. Asymmetric total power
gradients, smaller magnetic and optical pitch angles and a higher
polarization degree on one side of the disk were observed in the
wind-swept spirals NGC~2276 (Hummel
\&~Beck~\cite{hum95}) and NGC~4254 (Soida et
al.~\cite{soi96}, Chy\.zy et al. in prep.). NGC~2276 also has a
radio tail noted by Condon~(\cite{con83}). Pitch angle
asymmetries and an enhancement of the polarization on one side
of the disk by field compression are found in the tidally
interacting spiral NGC~3627 (Soida et al.~\cite{soi01} and
references therein). We note also that the large-scale
distribution of \ion{H}{i} presented by Thornley
\&~Mundy~(\cite{tho97}) shows at intermediate radii a
slightly larger gas extent and slower density decrease with
radius along the southern semi-axis, where we find larger
magnetic pitch angles. A similar coincidence is found in
interacting spirals (Chy\.zy et al. in prep.).

Isolated objects often show signs of past weak interactions, like
retrograde stellar orbits or \ion{H}{i} warps (e.g. Jore et
al.~\cite{jor96}). Multiple ``minor merging'' effects may occur
during the galaxy lifetime (Haynes et al.~\cite{hay00}),
sometimes producing large-scale asymmetries in spiral structure
(Zaritsky \&~Rix~\cite{zar97}) or short low-brightness stellar
or molecular tails (Aalto et al.~\cite{aal01}). Recently
Kornreich et al.~(\cite{kor01}) found that while the optical
morphology returns quickly to a symmetric appearance, the gas
dynamics constitutes a long-lasting memory of past, weak
perturbations.

NGC~4414 may be an example in which a distant interaction is
remembered by its magnetic field. The inner disk is relatively
unperturbed (see Sect. 4.2) and no central starburst nor a nuclear
source are present. NGC~4414 could have accreted diffuse gas,
perturbing only its outer disk. As demonstrated using numerical
models of galaxies interacting with ambient gas clouds
(Otmianowska-Mazur \&~Vollmer~\cite{otm02}), asymmetries
in the polarized intensity, magnetic pitch angles and \ion{H}{i}
may last for more than 1~Gyr. The magnetic field of NGC~4414
may thus ``remember'' a perturbation which could have occurred
1~Gyr ago. We note finally that the asymmetries found in the
H$\alpha$, \ion{H}{i} distribution and kinematics are very
weak and only visible in low-surface brightness features. They
are more obvious in the magnetic field properties like the pitch
angle variations by $45\degr$ and the dramatic asymmetry in
Faraday effects between the northern and southern disk.
Multifrequency polarization observations may thus serve as a
complementary tracer of weak perturbations, poorly visible in
other domains.

External influences could in principle explain the present-time
radial magnetic field component. However, a single event
cannot explain the observed strength of the regular magnetic
fields (see Sect. 4.1) which require stable, continuous
reproduction of their radial component over long timescales (e.g.
Brandenburg \&~Urpin~\cite{bra98}). Without constant
replenishment, the magnetic fields would diffuse out of the
galaxy in a timescale of 10$^{8}$ yrs (Parker~\cite{par79}).
Although one can still imagine a quasi-continuous set of
merging events which permanently regenerate the radial
magnetic field, the turbulent dynamo seems a more reasonable
alternative. {\it The observations presented here are evidence for
dynamo action with much less contamination from spiral arms
than in grand-design spirals, using a considerably better
resolution than in our earlier work (Knapik et
al.~\cite{kna00})}.

To reproduce the strength of the Faraday rotation is by far a
more complex issue. In the northern disk we got reasonable
agreement between our estimate of the regular magnetic field
and standard assumptions concerning the ionized gas properties.
We assume that about 80\% of the observed thermal flux
originates in \ion{H}{ii} regions making a small contribution to
the Faraday rotation because of their small volume filling factor.
Taking 20\% of the thermal flux to be due to a diffuse
component (Walterbos \&~Braun~\cite{wal94}) with a face-on
full thickness of 700~pc and a volume filling factor of 0.04
(Fletcher et al. in prep.) we obtain for the equipartition regular
field (assumed parallel to the disk plane) RM values of $\simeq
100$~rad/m$^2$, in reasonable agreement with observations. In
the southern disk the apparent RM jump signifies a Faraday-thick
regime and thus an intrinsic RM in excess of
500~rad/m$^2$. We can reproduce this taking the observed
thermal flux and assuming the same regular magnetic field but
assuming that 50\% (a factor 2.5 higher) of the observed thermal
emission comes from the diffuse gas. We also require a filling
factor of the diffuse ionized gas 2.5 times higher, equal to 0.1.
Additionally we need a much thicker magnetoionic disk, with a
face-on thickness of 3~kpc.

The observed magnetic field structure in NGC~4414 may be the
result of
the following processes:

\begin{itemize}

\item[-] The turbulent dynamo generates the global magnetic
field with a substantial radial component and a mixture of
modes.

\item[-] Past external interactions could enhance the regular field
in the northern half of the disk, decreasing the magnetic pitch
angles in this region and increasing them in the South, as found
in galaxies known to be affected by external influences
(Hummel \&~Beck~\cite{hum95}, Soida et al.~\cite{soi96},
Chy\.zy et al. in prep.). The non-axisymmetric fields may be
boosted by interactions (Moss~\cite{mos96b}) and/or from ram
pressure effects as modeled by Otmianowska-Mazur
\&~Vollmer~(\cite{otm02}).

\item[-] External interactions could thicken the magnetoionic
disk and make it contain more diffuse ionized gas in the South
than in the North. This could then make the southern disk of
NGC~4414 Faraday-thick at 4.86~GHz. Such a phenomenon is
so far unique among spiral galaxies.

\end{itemize}

We have found that the magnetic field may be a very sensitive
indicator of the behaviour of the diffuse component of the
interstellar gas. In particular, it may trace signs of interactions
which would pass unnoticed when observing the stellar or
H$\alpha$ emission. Studying their magnetic fields together
with deep mapping of their environment in \ion{H}{i}, X-rays
and in low-excitation spectral lines like \ion{C}{ii},
\ion{N}{ii}, \ion{O}{i} or \ion{O}{iii} provides an important
clue to their physics and evolution.

\section{Summary and conclusions}

We have performed a three-frequency VLA study of the
flocculent galaxy NGC~4414 known to have extremely weak
traces of optical spiral structure and no evidence for
non-axisymmetric gas flows. The data were analyzed together with
CO(1--0) and H$\alpha$ maps, yielding the following results:

\begin{itemize}

\item[-] The galaxy shows a bright total power disk with a
central depression corresponding to the minimum in H$\alpha$
and CO brightness, which may serve as an additional argument
for the absence of radial gas flows.

\item[-] The correlation of the nonthermal brightness with the
CO and H$\alpha$ emission also holds in a galaxy without
strong spiral arms and thus it is not due to an accumulation of
cold gas and star formation in density-wave compression
regions.

\item[-] Despite the lack of spiral arms and of non-azimuthal gas
flows the galaxy shows a clear magnetic spiral pattern with a
significant radial component, resembling that in grand-design
galaxies. Dynamo action is the most likely source of a
significant radial magnetic field component. Admixtures of
bisymmetric field and possibly higher modes are likely to exist
even in a flocculent galaxy.

\item[-] There is some correspondence between the optical
armlets discussed by Thornley \& Mundy (\cite{tho97}) and
local magnetic field orientations. The optical structure seems to
follow the general global asymmetry of magnetic pitch angles,
while the magnetic lines seem to be aligned with only
some segments of individual armlets.

\item[-] NGC~4414 shows a strongly asymmetric distribution of
Faraday rotation. Our observations are suggestive of a much
thicker magnetoionic disk and an increased relative content of
diffuse ionized gas (compared to classical \ion{H}{ii} regions)
in the southern disk. These asymmetries, together with those in
the magnetic pitch angles and total power asymmetries suggest
some past interaction (e.g. merging with dwarf galaxies or more
likely accreting gas clouds) in the history of NGC~4414.

\end{itemize}

In this work we show that the regular spiral magnetic pattern
observed in all nearby galaxies does not require strong density
wave action. We propose that the magnetic pattern in
NGC~4414 is due to the dynamo process free from
contamination by flows occurring in grand-design spiral arms.
On the other hand, the picture of the magnetic field in
NGC~4414 is still far from being clear and no good description
of the three-dimensional magnetic field geometry is yet
available, except for some rough and very qualitative guesses.
We suggest that observing the galactic magnetic field may
provide a clue to the properties of rarefied ionized gas,
contributing greatly to the Faraday rotation but little to the
H$\alpha$ emission. This gas phase may be more sensitive to
external interactions than either the molecular gas or
\ion{H}{ii} regions associated with recent star formation. Thus,
we propose to use the magnetic fields as an indicator of past
external interactions working even in cases when the optical
information does not indicate significant perturbations. We note
finally that a more quantitative magnetic field description in
NGC~4414 may come from extensive modeling using more
sophisticated multi-component field topologies and realistic
distributions of thermal gas and relativistic electrons. A deep
study of its environment in \ion{H}{i} line and X-ray is highly
desirable. Such a study is planned for the near future.

\begin{acknowledgements}
The Authors wish to express their thanks to Dr Richard W.
Pogge from Dept. of Astronomy, Ohio State University for
providing us with his H$\alpha$~map of the whole disk on
NGC~4414 in a numerical format and to Dr Wendy L.
Freedman from Carnegie Institution for her permission to use
the unpublished optical image. We express our thanks to Dr
Andrew Fletcher, University of Newcastle, for performing for us
the analysis of magnetic field modes in NGC~4414. We are
grateful to numerous colleagues from the Max-Planck-Institute
f\"ur Radioastronomie (MPIfR) in Bonn for their valuable
discussions during this work. M.U. and M.S. are indebted to
Professor R. Wielebinski from the MPIfR for the invitations to
stay at this Institute, where substantial parts of this work were
done. We wish to express our thanks to Professor Alexei Bykov
from Moscow State University for providing for us his program
DEPOLARM and his assistance in its use. We are also grateful
to colleagues from the Astronomical Observatory of the
Jagiellonian University in Krak\'ow for their comments. We
wish to express our gratitude to the anonymous referee for the
critical reading of our manuscript. This work was supported by a
grant from the Polish Research Committee (KBN), grants no.
962/P03/97/12 and 4264/P03/99/17.
\end{acknowledgements}

\end{document}